\documentclass[%
reprint,
amsmath,amssymb,
aps,
]{revtex4-2}

\usepackage{graphicx}% Include figure files
\usepackage{dcolumn}% Align table columns on decimal point
\usepackage{bm}% bold math
\usepackage{color}
\begin{document}

\preprint{APS/123-QED}

\title{Anomalous diffusion and transport by a reciprocal convective flow}

\author{Yuki Koyano}
\email{koyano@garnet.kobe-u.ac.jp}
\affiliation{Department of Human Environmental Science, Graduate School of Human Development and Environment, Kobe University, Kobe 657-0011, Japan}

\author{Hiroyuki Kitahata}
\email{kitahata@chiba-u.jp}
\affiliation{Department of Physics, Graduate School of Science, Chiba University, Chiba 263-8522, Japan
}

\date{\today}

\begin{abstract}
Under low-Reynolds-number conditions, dynamics of convection and diffusion are usually considered separately because their dominant spatial and temporal scales are different, but cooperative effects of convection and diffusion can cause diffusion enhancement [Koyano \textit{et al.}, \textit{Phys. Rev. E}, \textbf{102}, 033109 (2020)].
In this study, such cooperative effects are investigated in detail.
Numerical simulations based on the convection-diffusion equation revealed that anisotropic diffusion and net shift as well as diffusion enhancement occur under a reciprocal flow.
Such anomalous diffusion and transport are theoretically derived by the analyses of the Langevin dynamics.
\end{abstract}

\maketitle

\section{Introduction}

Mixing of fluid has been intensively studied not only from the scientific interest but also from the importance in the manufacturing industry.
On the mixing, two fundamental processes should be taken into account: the diffusion of the solute and the configuration changes of the fluid elements by the convective flow.
When the Reynolds number of the flow is high, the scales of these two processes are comparable and they are strongly coupled, which has been intensively studied as turbulent diffusion.~\cite{Thomson, Rodean}
In contrast, for the flow with low Reynolds number, the scales of these two processes are separated, i.e., the diffusion of the solute is the microscopic phenomenon, while the configuration changes of the fluid elements is macroscopic.
These two processes can be coupled despite the difference in scales, even in the case of the low Reynolds number~\cite{koyano2020}.
Here, we focus on the coupling of them under low-Reynolds-number conditions.

Phenomenologically, the dynamics of a concentration field of solute molecules under flows is often described by the convection-diffusion equation, which can be derived as the Fokker-Planck equation from the Langevin equation including convective effect~\cite{Risken}.
In the diffusion-convection equation, the convection and diffusion terms appear separately, reflecting that these processes are independent.
When we consider the finite-time evolution of the convection-diffusion equation, however, the time evolution of the concentration field is not obtained by simply adding the evolutions due to the convection and diffusion processes, but the coupling of these processes should also be included.

Actually, there are several systems in which the cooperative dynamics of convection and diffusion play an essential role.
For example, the coupling of convection and diffusion is useful to mix two fluids efficiently in narrow microfluidic channels~\cite{CEJ, MolSci, Nady}, since it is difficult to induce turbulence in such channels due to the low Reynolds number.
For the same reason, the coupling of them is also useful for mixing in micro-droplet reactors like levitated small droplets by the periodically modulated electric or acoustic field~\cite{Watanabe, Hasegawa}, small tanks vibrated by acoustic wave~\cite{Shilton}, and merged droplets by gas flow~\cite{Carroll}.
Moreover, there have been several reports that the diffusion is enhanced inside cells~\cite{Cell, Parry}, and such diffusion enhancement can be understood from the coupling of the thermal diffusion and flow induced by the actively moving proteins or cell deformation~\cite{PNAS, PhysicaD, PRE, JPSJ, Xu}.

\begin{figure}
	\centering
	\includegraphics{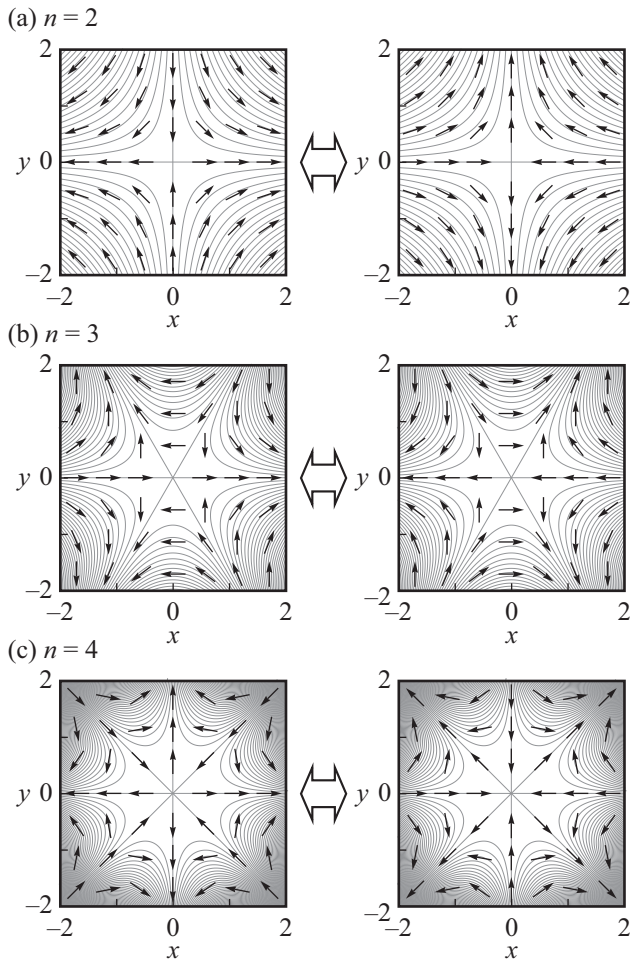}
	\caption{Profile of the reciprocal flow expressed in Eqs.~\eqref{reciprocal_flow} and \eqref{def_Psi}.
	The direction of the flow is inverted periodically.
	The gray curves show streamlines, while arrows show the direction of the flow.
	The modes of the reciprocal flow are (a) $n=2$, (b) $n=3$, (c) $n=4$.}
	\label{fig_added}
\end{figure}

In the previous paper~\cite{koyano2020}, we derived the discretized Fokker-Planck equation, and formulated the coupling of the convection and diffusion processes for a given time interval, especially in the case where a reciprocal flow is considered.
Here, the reciprocal flow is defined as the oscillatory flow with time reversal symmetry.
In other words, all the fluid elements move back and forth, and there is no circulation (cf.~Fig.~\ref{fig_added}).
The net convection by the reciprocal flow is zero everywhere after a period of the flow.
By calculating the diffusion term in the discretized Fokker-Planck equation, it is found that the effective diffusion coefficient, which is defined by the mean square displacement divided by the finite time interval, is larger than the thermal diffusion coefficient.
This means that the coupling of the convection and diffusion results in the diffusion enhancement.

\begin{figure*}
	\centering
	\includegraphics{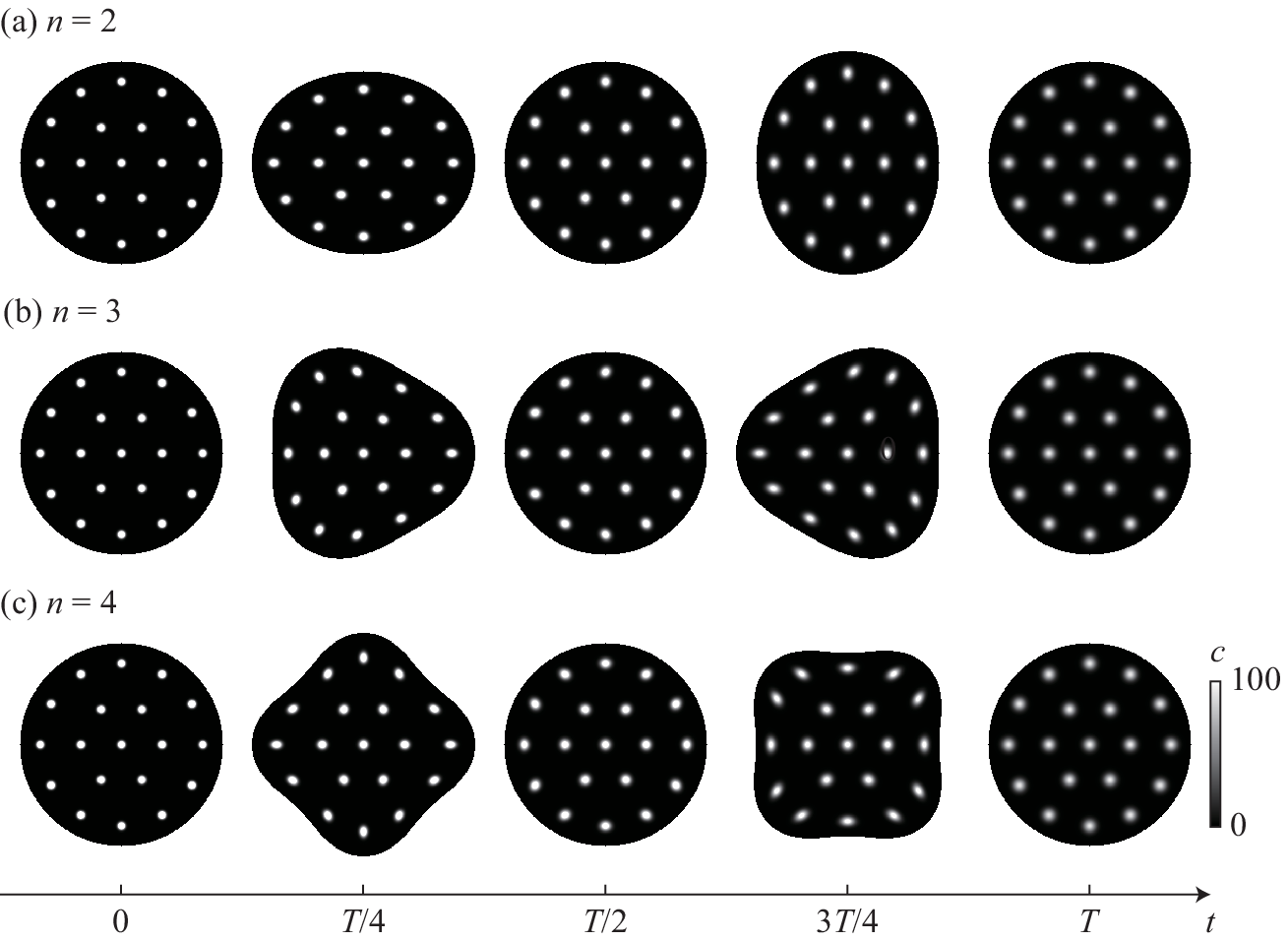}
	\caption{Time evolutions of the concentration profile $c$ during one period $T$ of the reciprocal flow, which are numerically calculated based on Eq.~\eqref{cd_eq}.
	The white spots indicate the high-concentration areas, which are convected and deformed by the reciprocal flow.
	The modes of the reciprocal flow are (a) $n = 2$, (b) $n = 3$, and (c) $n = 4$.
	The black region indicates the area occupied by fluid elements in a circular region with a radius of 1 at $t = 0$ to show the configuration changes of fluid elements.
	The thermal diffusion coefficient $D$ and the amplitude of the flow $\varepsilon$ were set to be $D = 10^{-4}$ and $\varepsilon = 10^{-1}$, respectively.}
	\label{fig1}
\end{figure*}

In this study, we analyzed the cooperative dynamics of convection and diffusion more in detail.
We first show the numerical results on the time evolution of diffusion process under a reciprocal flow.
We find the three types of dynamics, a diffusion enhancement, an anisotropic diffusion, and a net shift.
As we mentioned in the previous paragraph, the first one has been already reported~\cite{koyano2020}, but the others are newly found.
As for the anomalous diffusion, i.e., the diffusion enhancement and the anisotropic diffusion, either of which is dominant depends on the ratio between the amplitude of the reciprocal flow and the diffusion coefficient originating from the thermal fluctuations.
We also numerically observed the shift in the center position (the first moment) of the concentration profile after one period of a reciprocal flow.
Such shift is nontrivial, since the fluid element does not shift at all after one period of the reciprocal flow.
To understand the diffusion enhancement, anisotropic diffusion, and net shift, the Langevin dynamics of a solute under a reciprocal flow is analyzed.

The present paper is organized as follows: First, the numerical results based on the convection-diffusion equation are provided in Sec.~II.
In Sec.~III, the Langevin dynamics is theoretically reduced into the discretized Fokker-Planck equation.
In Sec.~IV, the theoretical results are compared with numerical ones to confirm that the three types of dynamics are well reproduced by the theoretical analysis.

\section{Numerical results}

In the present paper, we consider a two-dimensional system.
First, the numerical results of the convection-diffusion equation:
\begin{align}
\frac{\partial c(\bm{x},t)}{\partial t} = - \bm{v}(\bm{x},t) \cdot \nabla c(\bm{x},t) + D\nabla^2 c(\bm{x},t) \label{cd_eq}
\end{align}
are shown.
Here, $c$ is a solute concentration, $\bm{v}$ is a flow field, and $D$ is a thermal diffusion coefficient.
As for the flow field $\bm{v}$, we consider the reciprocal flow field:
\begin{align}
\bm{v}(\bm{x},t) = \nabla \Psi(\bm{x}) \cos \omega t, \label{reciprocal_flow}
\end{align}
which is factorized into the space- and time-dependent parts.
Note that $\bm{v}$ satisfies the incompressibility condition.
$\omega$ is the angular frequency of the reciprocal flow, and is set to be 1.
As for the function $\Psi$, we consider the $n$-th mode flow $\Psi(\bm{x}) = \Psi^{(n)} (\bm{x})$, which is defined as:
\begin{align}
\Psi^{(n)} (\bm{x}) = \frac{\varepsilon}{n} r^n \cos n \theta, \label{def_Psi}
\end{align}
where %the index $n$ is the mode of flow field, 
$\varepsilon$ is the amplitude of the reciprocal flow and the polar coordinates $\bm{x} = r (\cos \theta \bm{e}_x + \sin\theta \bm{e}_y)$ are adopted.
The flow profiles are exemplified in Fig.~\ref{fig_added}.
The detailed derivation of the flow field $\bm{v}$ is shown in Appendix~\ref{FF}.

First, we investigated the change of a spot-like profile after a period of the reciprocal flow.
To observe the dependence of the initial position of the spot, we considered multiple spots as an initial condition, which is given by the summation of several Gaussian distributions as
\begin{align}
c(\bm{x},0) = \sum_{i} \frac{1}{2\pi \sigma^2} \exp \left( - \frac{\left| \bm{x} - \bm{x}_{\mathrm{ini},i} \right|^2}{2\sigma^2} \right), \label{ini_cond_multi}
\end{align}
where $\sigma$ is a characteristic size of the localized distribution.
The distribution of $\bm{x}_{\mathrm{ini},i}$ is illustrated by white spots on the snapshot at $t=0$ in Fig.~\ref{fig1}.
Here, we set $\sigma = 0.02$, and thus each Gaussian distribution in the initial condition approximates the Dirac's delta function.
The Dilichlet boundary condition with $c = 0$ is set at a circle with a radius of 2 as the boundary.
The numerical calculations are performed by the finite-difference discretization in space, and explicit (Euler) discretization with the spatial mesh of $5 \times 10^{-3}$ and the time step of $2\pi \times 10^{-5}$.

The time series of the concentration profile $c$ during a period of the reciprocal flow $T = 2\pi/\omega$ are displayed in Fig.~\ref{fig1}.
The white spots where the concentration was higher were convected and deformed by the reciprocal flow, but seemed to be back to the initial position after a period $T$.
This is because we adopted the reciprocal flow as a convectional flow.
In addition, they were blurred a little, which indicates the high-concentration regions were expanded and the peak values are decreased by the diffusion.

\begin{figure}
	\centering
	\includegraphics{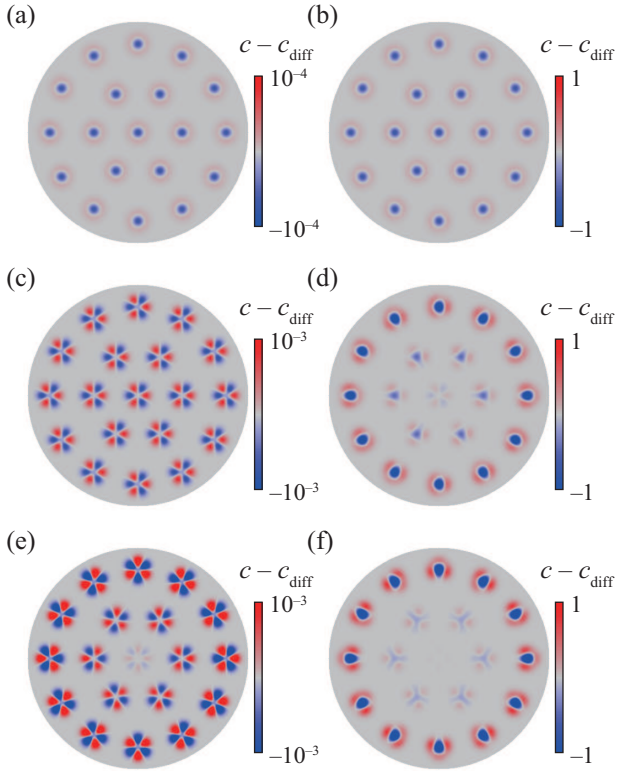}
	\caption{Snapshots of the concentration difference $c(\bm{x},T) - c_\mathrm{diff} (\bm{x},T)$.
	The modes of the reciprocal flow are (a,b) $n = 2$, (c,d) $n = 3$, and (e,f) $n = 4$.
	The red and blue regions show positive and negative values, respectively.
	The thermal diffusion coefficient was set as $D= 10^{-4}$ and the amplitude of the reciprocal flow was set as (a,c,e) $\varepsilon = 10^{-3}$ and (b,d,f) $\varepsilon = 10^{-1}$. The gray region indicates a circular region with a radius of 1.
	}
	\label{fig2}
\end{figure}

To clarify the cooperative effect of the convection and diffusion, we also calculated the dynamics of concentration field without any convection, i.e., the diffusion dynamics:
\begin{align}
\frac{\partial c_\mathrm{diff} (\bm{x},t)}{\partial t} = D\nabla^2 c_\mathrm{diff} (\bm{x},t)
\end{align}
with the initial condition in Eq.~\eqref{ini_cond_multi}, and used $c_\mathrm{diff}$ as a reference.
The snapshots of $c(\bm{x},T) - c_\mathrm{diff} (\bm{x},T)$ are shown in Fig.~\ref{fig2} for the reciprocal flow with $n = 2$, $3$, and $4$.
We can see that the profiles of $c(\bm{x},T)$ differ from those of $c_\mathrm{diff} (\bm{x},T)$.
In the case of $n=2$, the concentration difference became higher and lower at outer and inner regions around each $\bm{x}_{\mathrm{ini},i}$, respectively.
Such a type of pattern indicates the diffusion was enhanced by the reciprocal flow.
In the cases of $n=3$ and $4$, the three-pronged patterns were observed for $\varepsilon = 10^{-3}$, while diffusion enhancement was observed for $\varepsilon = 10^{-1}$.
The three-pronged pattern indicates that the diffusion is anisotropic.

\begin{figure}
	\centering
	\includegraphics{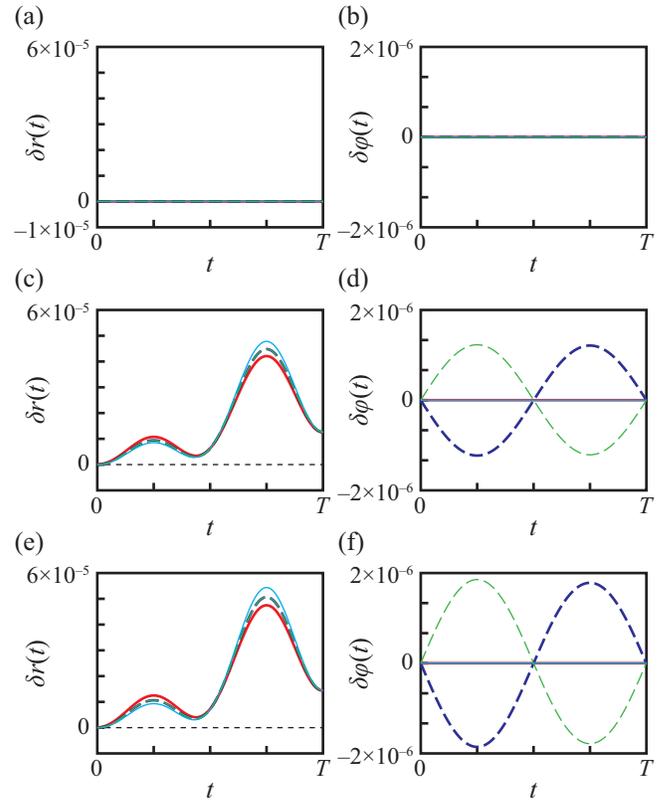}
	\caption{Time series of (a,c,e) $\delta r(t)$ and (b,d,f) $\delta \varphi (t)$.
	The initial conditions were set to be $r_\mathrm{ini} = 0.5$ and $\theta_\mathrm{ini}= 0$ (red solid curve), $\pi/(2n)$ (light green broken curve), $\pi/n$ (light blue solid curve), and $3\pi/(2n)$ (dark blue broken curve) for the $n$-th mode flow ((a,b) $n = 2$, (c,d) $n = 3$, and (e,f) $n = 4$).
	Other parameters were set to be $\varepsilon = 0.1$ and $D = 10^{-4}$.
	}
	\label{fig3}
\end{figure}

\begin{figure}
	\centering
	\includegraphics{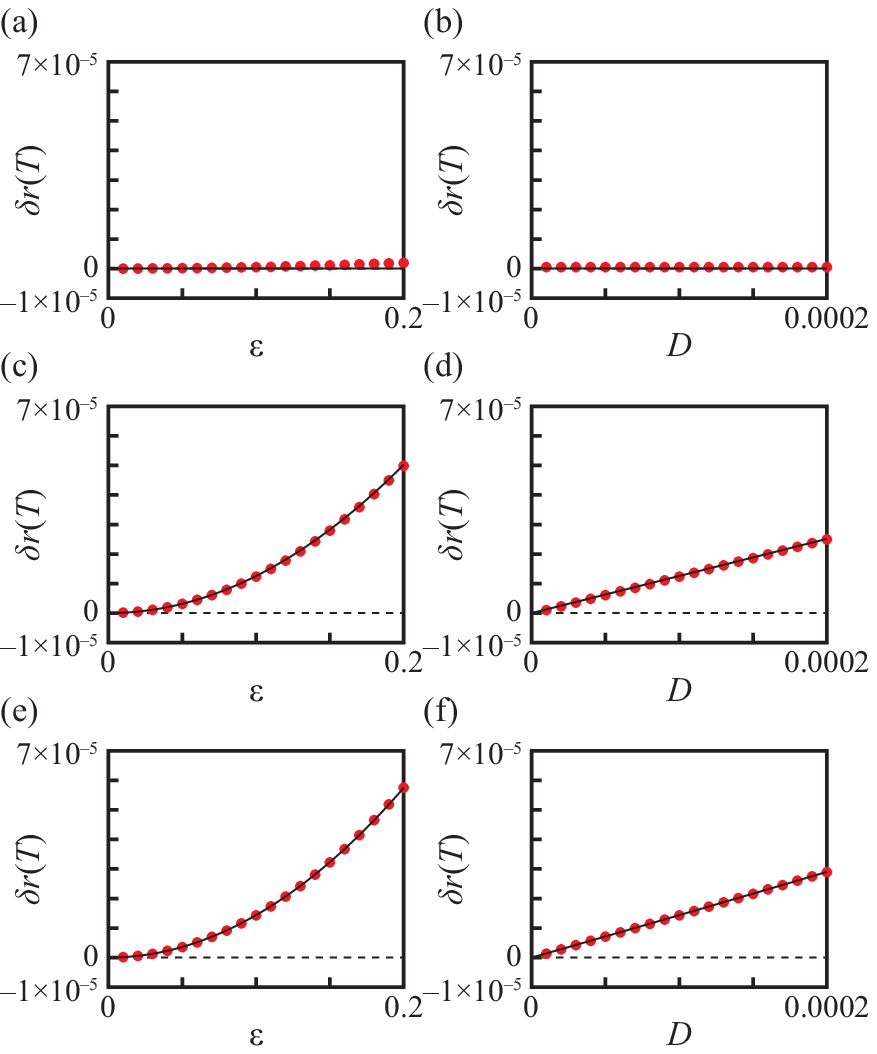}
	\caption{Plots of $\delta r(T)$ against (a,c,e) the flow amplitude $\varepsilon$ and (b,d,f) the diffusion coefficient $D$.
	The initial conditions were $r_\mathrm{ini} = 0.5$ and $\theta_\mathrm{ini}= 0$ for the $n$-th mode ((a,b) $n = 2$, (c,d) $n = 3$, and (e,f) $n = 4$).
	The parameters were fixed as $D=10^{-4}$ for (a,c,e) and $\varepsilon = 10^{-1}$ for (b,d,f).
The initial size of the spot was fixed for $\sigma = 0.02$.
	The black solid curves show the theoretical estimation, which is given in Eq.~\eqref{net_shift}.}
	\label{fig4}
\end{figure}

Next, we numerically investigate the dynamics of the center position of the concentration profile.
In this case, we adopt a single Gaussian distribution:
\begin{align}
c(\bm{x},0) = \frac{1}{2\pi \sigma^2} \exp \left( - \frac{\left| \bm{x} - \bm{x}_{\mathrm{ini}} \right|^2}{2\sigma^2} \right) \label{c_ini}
\end{align}
as the initial condition instead of Eq.~\eqref{ini_cond_multi}.
The numerical methods, the boundary condition, and the spatial and temporal meshes are the same as those in the case of the multiple spots.
The center position (the first moment) of the concentration profile is defined as 
\begin{align}
\overline{\bm{x}}(t) = \frac{\int \bm{x} c(\bm{x}, t) d\bm{x}}{\int c(\bm{x}, t) d\bm{x}}, \label{pathline}
\end{align}
and the time development of $\overline{\bm{x}}(t)$ during the period of the reciprocal flow is observed.
It should be noted that $\overline{\bm{x}}(0) = \bm{x}_\mathrm{ini}$.
Though the reciprocal flow does not cause net drift after a period, the drift by the convection takes place during the period, which is included in Eq.~\eqref{pathline}.
To clarify the effect of the cooperative effect of diffusion and convection, the effect of pure convection should be subtracted. Thus, we define $\delta \bm{r}(t)$ as:
\begin{align}
\delta \bm{r}(t) = \overline{\bm{x}}(t) - \bm{r} (t; \bm{x}_\mathrm{ini}),
\end{align}
where $\bm{r}(t; \bm{x}_\mathrm{ini})$ is the pathline of a fluid element that is initially located at $\bm{x}_\mathrm{ini}$, i.e.,
\begin{align}
\bm{r}(t; \bm{x}_\mathrm{ini}) = \bm{x}_\mathrm{ini} + \int_0^t \bm{v}(\bm{r}(t'; \bm{x}_\mathrm{ini}), t') dt'.
\end{align}

In Fig.~\ref{fig3}, the time series of $\delta r(t)$ and $\delta \varphi(t)$ are plotted.
Here, we defined $\delta r(t)$ as $\delta r(t) = |\delta \bm{r}(t)|$ and $\delta \varphi (t)$ as the signed angle between $\bm{r}(t; \bm{x}_\mathrm{ini})$ and $\overline{\bm{x}}(t)$.
In the case of $n=2$, $\delta \bm{r}(t)$ was always zero.
On the other hand, $\delta \bm{r}(t)$ changed in the cases of $n = 3$ and $4$.
Since $\delta \bm{r}$ is the deviation of the center position of the spot from the position of the fluid element which locates at the center position of the initial spot, $\delta \bm{r}$ should be always zero if the cooperative effect of diffusion and convection is absent.
Thus, there is a shift of the center position of the spot caused by the cooperative effect of diffusion and convection for $n=3$ and 4.
In Fig.~\ref{fig4}, $\delta r(T)$ is plotted against $\varepsilon$ and $D$.
In the case of $n=2$, $\delta r(T)$ was always zero.
In the cases of $n=3$ and 4, $\delta r(T)$ was finite for the case that the flow amplitude $\varepsilon$ and the thermal diffusion coefficient $D$ are both non-zero.
These results indicate that the extra shift occurs due to the cooperative effect of convection and diffusion.

\section{Theoretical analysis\label{sec_theory}}

In this section, the diffusion enhancement, anisotropic diffusion, and net shift originating from the cooperative effect of convection and diffusion are theoretically explained.
We consider the overdamped Langevin dynamics with the reciprocal flow $\bm{v}(\bm{x}(t),t)$:
\begin{align}
\frac{d\bm{x}}{dt} = \bm{v}(\bm{x}(t),t) + \bm{\xi}(t), \label{Langevin}
\end{align}
where the reciprocal flow $\bm{v}$ was defined in Eq.~\eqref{reciprocal_flow}.
The function $\bm{\xi}(t)$ represents the thermal noise, which satisfies the following relations:
\begin{align}
\left < \bm{\xi}(t) \right > &= \bm{0}, \\
\left < \xi_\alpha (t) \xi_\beta (s) \right > &= 2 D \delta_{\alpha\beta} \delta(t-s),
\end{align}
where $\delta_{\alpha\beta}$ is the Kronecker delta, and $\left < \cdots \right >$ means the ensemble average.

The Fokker-Planck equation corresponding to Eq.~\eqref{Langevin} is obtained as 
\begin{align}
\frac{\partial q(\bm{x},t)}{\partial t} = - \nabla \cdot \left ( \bm{v} (\bm{x},t) q(\bm{x},t) \right ) + D \nabla^2 q(\bm{x},t), \label{FP}
\end{align}
where $q(\bm{x},t)$ is the probability density of $\bm{x}(t)$.
To discuss the diffusion enhancement, anisotropic diffusion, and net shift, we can investigate the profile of $q$ by integrating Eq.~\eqref{FP} over one period of the reciprocal flow.

Another approach is to derive a ``discrete Fokker-Planck equation''~\cite{koyano2020}:
\begin{align}
&q(\bm{x},T) - q(\bm{x},0) \nonumber \\
&= \sum_{m=1}^{\infty} \frac{(-1)^m}{m!} \frac{\partial^m}{\partial x_{\alpha_1} \cdots \partial x_{\alpha_m}} \left ( M^{(m)}_{\alpha_1 \cdots \alpha_m}(\bm{x},T) \; q(\bm{x},0) \right ), \label{dFPE}
\end{align}
where the dynamics during one period $T$ is included in coefficients $M_{\alpha_1 \cdots \alpha_m}(\bm{x},T)$.
Here and below, summation over repeated Greek indices is always assumed.
$M_{\alpha_1 \cdots \alpha_m}(\bm{x},T)$ is the $m$-th moment of $\bm{x}$ for the time interval of $T$:
\begin{align}
M^{(m)}_{\alpha_1 \cdots \alpha_m}(\bm{x},T) = \left < \prod_{i=1}^m \left [ x_{\alpha_i}(T) - x_{\alpha_i}(0) \right ] \right >.
\end{align}
Here, we calculate them up to the fourth order, and denote as $X_\alpha = M^{(1)}_{\alpha}$, $S_{\alpha\beta} = M^{(2)}_{\alpha\beta}$, $Y_{\alpha\beta\gamma} = M^{(3)}_{\alpha\beta\gamma}$, and $K_{\alpha\beta\gamma\mu} = M^{(4)}_{\alpha\beta\gamma\mu}$.
We have the following expressions for the reciprocal flow in Eq.~\eqref{reciprocal_flow}:
\begin{widetext}
\begin{align}
X_\alpha (\bm{x}, T) 
&= \frac{D\varepsilon^2 T}{2\omega^2} \frac{\partial^3 \Psi (\bm{x})}{\partial x_\alpha \partial x_{\alpha'} \partial x_{\alpha''}} \frac{\partial^2 \Psi (\bm{x})}{\partial x_{\alpha'} \partial x_{\alpha''}}
+ \frac{D^2\varepsilon^2 T^2}{2\omega^2} \frac{\partial^4 \Psi (\bm{x})}{\partial x_\alpha \partial x_{\alpha'} \partial x_{\alpha''} \partial x_{\alpha'''}} \frac{\partial^3 \Psi (\bm{x})}{\partial x_{\alpha'} \partial x_{\alpha''} \partial x_{\alpha'''}}
+ \mathcal{O} \left ( \varepsilon^3, D^3 \right ),
\end{align}
\begin{align}
S_{\alpha \beta} (\bm{x}, T) &= 2DT \delta_{\alpha \beta}
+ \frac{2 D\varepsilon^2 T}{\omega^2} \frac{\partial^2 \Psi (\bm{x})}{\partial^2 x_\alpha \partial x_{\alpha'}} \frac{\partial^2 \Psi (\bm{x})}{\partial x_\beta \partial x_{\alpha'}}
- \frac{D\varepsilon^2 T}{\omega^2} \frac{\partial^3 \Psi (\bm{x})}{\partial x_\alpha \partial x_\beta \partial x_{\alpha'} }\frac{\partial \Psi (\bm{x})}{\partial x_{\alpha'}} \nonumber \\
&\quad + \frac{3D^2 \varepsilon^2 T^2}{\omega^2} \frac{\partial^3 \Psi (\bm{x})}{\partial x_\alpha \partial x_{\alpha'} \partial x_{\alpha''}} \frac{\partial^3 \Psi (\bm{x})}{\partial x_\beta \partial x_{\alpha'} \partial x_{\alpha''}} 
+ \mathcal{O} \left ( \varepsilon^3, D^3 \right ),
\end{align}
\begin{align}
Y_{\alpha \beta \gamma} (\bm{x}, T) 
&= \frac{24 D^2 \varepsilon T}{\omega^2} \frac{\partial^3 \Psi (\bm{x})}{\partial x_\alpha \partial x_\beta \partial x_\gamma}
- \frac{3 D^2 \varepsilon^2 T^2}{\omega^2} \frac{\partial^4 \Psi (\bm{x})}{\partial x_\alpha \partial x_\beta \partial x_\gamma \partial x_{\alpha'}} \frac{\partial \Psi (\bm{x})}{\partial x_{\alpha'}} \nonumber \\
&\quad + \frac{3 D^2 \varepsilon^2 T^2}{\omega^2} \left (
\frac{\partial^3 \Psi (\bm{x})}{\partial x_\alpha \partial x_\beta \partial x_{\alpha'}} \frac{\partial^2 \Psi (\bm{x})}{\partial x_\gamma \partial x_{\alpha'}} 
+\frac{\partial^3 \Psi (\bm{x})}{\partial x_\beta \partial x_\gamma \partial x_{\alpha'}}
\frac{\partial^2 \Psi (\bm{x})}{\partial x_\alpha \partial x_{\alpha'}} + \frac{\partial^3 \Psi (\bm{x})}{\partial x_\gamma \partial x_\alpha \partial x_{\alpha'}} \frac{\partial^2 \Psi (\bm{x})}{\partial x_\beta \partial x_{\alpha'}} 
\right ) \nonumber \\
&\quad +\frac{D^2 \varepsilon^2 T^2}{\omega^2} \left ( \frac{\partial^3 \Psi (\bm{x})}{\partial x_\alpha \partial x_{\alpha'} \partial x_{\alpha''}} \frac{\partial^2 \Psi (\bm{x})}{\partial x_{\alpha'} \partial x_{\alpha''}} \delta_{\beta \gamma}
+ \frac{\partial^3 \Psi (\bm{x})}{\partial x_\beta \partial x_{\alpha'} \partial x_{\alpha''}} \frac{\partial^2 \Psi (\bm{x})}{\partial x_{\alpha'} \partial x_{\alpha''}} \delta_{\gamma \alpha}
+ \frac{\partial^3 \Psi (\bm{x})}{\partial x_\gamma \partial x_{\alpha'} \partial x_{\alpha''}} \frac{\partial^2 \Psi (\bm{x})}{\partial x_{\alpha'} \partial x_{\alpha''}} \delta_{\alpha \beta} \right ) \nonumber \\
&\quad + \mathcal{O} (\varepsilon^3, D^3),
\end{align}
\begin{align}
K_{\alpha \beta \gamma \mu} (\bm{x}, T)
&= 4D^2 T^2 (\delta_{\alpha\beta} \delta_{\gamma\mu} + \delta_{\alpha\gamma} \delta_{\beta\mu} + \delta_{\alpha\mu} \delta_{\beta\gamma}) \nonumber \\
&\quad + \frac{4D^2 \varepsilon^2 T^2}{\omega^2} \left ( \frac{\partial^2 \Psi (\bm{x})}{\partial x_\alpha \partial x_{\alpha'}} \frac{\partial^2 \Psi (\bm{x})}{\partial x_\beta \partial x_{\alpha'}} \delta_{\gamma \mu}
+ \frac{\partial^2 \Psi (\bm{x})}{\partial x_\alpha \partial x_{\alpha'}} \frac{\partial^2 \Psi (\bm{x})}{\partial x_\gamma \partial x_{\alpha'}} \delta_{\beta \mu}
+\frac{\partial^2 \Psi (\bm{x})}{\partial x_\alpha \partial x_{\alpha'}} \frac{\partial^2 \Psi (\bm{x})}{\partial x_\mu \partial x_{\alpha'}} \delta_{\beta \gamma} \right . \nonumber \\
& \qquad \qquad \qquad \left . 
+ \frac{\partial^2 \Psi (\bm{x})}{\partial x_\beta \partial x_{\alpha'}} \frac{\partial^2 \Psi (\bm{x})}{\partial x_\gamma \partial x_{\alpha'}} \delta_{\alpha \mu}
+ \frac{\partial^2 \Psi (\bm{x})}{\partial x_\beta \partial x_{\alpha'}} \frac{\partial^2 \Psi (\bm{x})}{\partial x_\mu \partial x_{\alpha'}} \delta_{\alpha \gamma}
+\frac{\partial^2 \Psi (\bm{x})}{\partial x_\gamma \partial x_{\alpha'}} \frac{\partial^2 \Psi (\bm{x})}{\partial x_\mu \partial x_{\alpha'}} \delta_{\alpha \beta} \right ) \nonumber \\
&\quad - \frac{2D^2 \varepsilon^2 T^2}{\omega^2} \left ( 
\frac{\partial^3 \Psi (\bm{x})}{\partial x_\alpha \partial x_\beta \partial x_{\alpha'}} \delta_{\gamma \mu} 
+ \frac{\partial^3 \Psi (\bm{x})}{\partial x_\alpha \partial x_\gamma \partial x_{\alpha'}} \delta_{\beta \mu} 
+ \frac{\partial^3 \Psi (\bm{x})}{\partial x_\alpha \partial x_\mu \partial x_{\alpha'}} \delta_{\beta \gamma} \right . \nonumber \\
& \qquad \qquad \qquad \left .
+ \frac{\partial^3 \Psi (\bm{x})}{\partial x_\beta \partial x_\gamma \partial x_{\alpha'}} \delta_{\alpha \mu} 
+ \frac{\partial^3 \Psi (\bm{x})}{\partial x_\beta \partial x_\mu \partial x_{\alpha'}} \delta_{\alpha \gamma} 
+ \frac{\partial^3 \Psi (\bm{x})}{\partial x_\gamma \partial x_\mu \partial x_{\alpha'}} \delta_{\alpha \beta} 
\right ) \frac{\partial \Psi (\bm{x})}{\partial x_{\alpha'}} + \mathcal{O} (\varepsilon^3, D^3).
\end{align}
\end{widetext}
The above moments were calculated up to the second order of $D$ and $\varepsilon$, and numerically confirmed as shown in Appendix~\ref{Confirm}.
The fifth and higher moments are in the order of $D^3$ or $\varepsilon^3$, and thus they are neglected hereafter.

\section{Discussion}

In this section, the analytical results are discussed through the comparison with the numerical results.
Each term in $X_\alpha$, $S_{\alpha\beta}$, $Y_{\alpha\beta\gamma}$, and $K_{\alpha\beta\gamma\mu}$ except $2DT\delta_{\alpha\beta}$ in $S_{\alpha\beta}$ and $4D^2T^2(\delta_{\alpha\beta} \delta_{\gamma\mu} + \delta_{\alpha\gamma} \delta_{\beta\mu} + \delta_{\alpha\mu} \delta_{\beta\gamma})$ in $K_{\alpha\beta\gamma\mu}$ includes both the amplitude of reciprocal flow $\varepsilon$ and the diffusion coefficient $D$ originating from the thermal noise.
Such terms show the cooperative effect of convection and diffusion.

The effective diffusion coefficient $D^\mathrm{eff}$ is defined as the mean square displacement divided by $4T$, and thus is explicitly described as
\begin{align}
D^\mathrm{eff} =& \frac{1}{4T} S_{\alpha \alpha} \\
=& D + (n-1)^2 \frac{D\varepsilon^2}{\omega^2} r^{2(n-2)} \nonumber \\
& + 3 (n-1)^2 (n-2)^2 \frac{D^2 \varepsilon^2 T}{\omega^2} r^{2(n-3)}, \label{diff_enhance}
\end{align}
where $T$ is the period of the reciprocal flow, and $(r, \theta)$ is the position in the polar coordinates.
Obviously, $D^\mathrm{eff}$ is larger than $D$.
In other words, the diffusion is enhanced by the reciprocal flow.
For example, the effective diffusion coefficients $D^\mathrm{eff,2}$, $D^\mathrm{eff,3}$, and $D^\mathrm{eff,4}$ for the reciprocal flow with $n = 2$, 3, and 4 are calculated as
\begin{align}
D^\mathrm{eff,2} =& D + \frac{D \varepsilon^2}{\omega^2}, \\
D^\mathrm{eff,3} =& D + \frac{8D \varepsilon^2}{\omega^2} r^2 + \frac{12 D^2 \varepsilon^2 T}{\omega^2}, \\
D^\mathrm{eff,4} =& D + \frac{9 D \varepsilon^2}{\omega^2} r^4 + \frac{108 D^2 \varepsilon^2 T}{\omega^2} r^2.
\end{align}
The diffusion is uniformly enhanced for $n=2$, while the diffusion enhancement has a spatial dependence for $n=3$ and 4.

\begin{figure}
	\centering
	\includegraphics{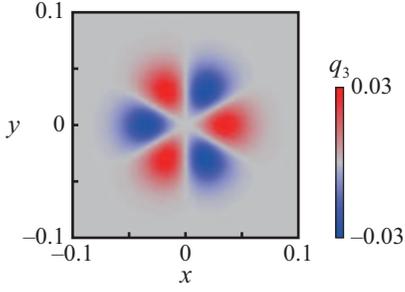}
	\caption{Profile of $q_3 (\bm{x})$ in Eq.~\eqref{anisotropic}. The parameters were set to be $D=10^{-4}$, $\varepsilon=10^{-3}$, and $\sigma = 0.02$.}
	\label{fig5}
\end{figure}

Then, the effect of the third moments is discussed.
The explicit forms of the third moments are described as follows:
\begin{align}
Y_{xxx} =& - Y_{xyy} \nonumber \\
	=& 24 (n-1) (n-2) \frac{D^2 \varepsilon T}{\omega^2} r^{n-3} \cos (n-3) \theta, \label{anisotropic_diff_1}\\
Y_{yyy} =& - Y_{xxy} \nonumber \\
	=& 24 (n-1) (n-2) \frac{D^2 \varepsilon T}{\omega^2} r^{n-3} \sin (n-3) \theta. \label{anisotropic_diff_2}
\end{align}
Here, we exemplify the case for $n=3$.
The third moments for $n=3$ are
\begin{align}
Y_{xxx} =& - Y_{xyy} = \frac{48 D^2 \varepsilon T}{\omega^2}, \\
Y_{yyy} =& - Y_{xxy} = 0.
\end{align}
As the concentration profile at $t=0$, we adopt a single spot as in Eq.~\eqref{c_ini} with $\bm{x}_\mathrm{ini} = \bm{0}$, and then the modulation of concentration field by the third moments, $q_3(\bm{x})$, is calculated as follows:
\begin{align}
q_3 (\bm{x})
&= - \frac{1}{6} \frac{\partial^3}{\partial x_\alpha \partial x_\beta \partial x_\gamma} \left ( Y_{\alpha \beta \gamma} q(\bm{x},0) \right ) \nonumber \\
&= \frac{8 D^2 \varepsilon}{\omega^3 \sigma^8} \left ( x^3 - 3 x y^2 \right ) \exp \left ( - \frac{|\bm{x}|^2}{2\sigma^2} \right ). \label{anisotropic}
\end{align}
The profile of Eq.~\eqref{anisotropic} is displayed in Fig.~\ref{fig5}.
The three-pronged pattern means that the concentration profile becomes anisotropic with regard to $\bm{x} = \bm{0}$ after one period.
This suggests that the third-order term is the origin of the anisotropic transport.

As numerical simulations and theoretical analysis show, the anomalous diffusion occurs as a result of the cooperative effect of the convection and diffusion.
The leading term of the diffusion enhancement is on the order of $D \varepsilon^2$ as in Eq.~\eqref{diff_enhance}, while that of the anisotropic diffusion is $D^2 \varepsilon$ as in Eqs.~\eqref{anisotropic_diff_1} and \eqref{anisotropic_diff_2}.
This indicates that the diffusion enhancement (anisotropic diffusion) is dominant for larger (smaller) flow amplitude and smaller (larger) thermal diffusion coefficient, which can be clearly seen in Fig.~\ref{fig2}.
It is noted that the third moments are 0 for $n=2$, and thus the anisotropic diffusion does not appear in Fig.~\ref{fig2}(a,b).

For the reciprocal flow expressed in Eq.~\eqref{def_Psi}, the net shift $X_\alpha$ is calculated as 
\begin{align}
X_r =& \frac{D\varepsilon^2 T}{\omega^2} (n-1)^2 (n-2) r^{2n-5} \nonumber \\
& + \frac{2D^2\varepsilon^2T^2}{\omega^2} (n-1)^2 (n-2) (n-3) r^{2n-7}, \label{xr}\\
X_\theta =& 0, \label{xtheta}
\end{align}
where $X_r$ and $X_\theta$ are the radial and angular components of $X$.
Equations \eqref{xr} and \eqref{xtheta} indicate that the net shift is in the outward direction from the origin.
The explicit forms of $X_r$ for $n=2$, 3, and 4 are described as follows:
\begin{align}
X^{2}_r =& 0, \\
X^{3}_r =& \frac{4 D \varepsilon^2 T}{\omega^2} r, \\
X^{4}_r =& \frac{18 D \varepsilon^2 T}{\omega^2} r^3 + \frac{72 D^2 \varepsilon^2 T^2}{\omega^2} r.
\end{align}
For $n=2$, the net shift is zero.
For $n=3$ and 4, the net shift becomes larger for larger $r$, which reflects the strong spatial dependence of the reciprocal flow.
The above expression of the net shift is for a single particle, i.e., the case that the initial concentration profile is given by the Dirac's delta function.
The general expression of the net shift after a period of the reciprocal flow is given by
\begin{align}
\delta r_\alpha = \int X_\alpha(\bm{r}') c(\bm{r}',0) d\bm{r}', \label{net_shift_gen}
\end{align}
where $c(\bm{r},0)$ is the initial profile.
When $c(\bm{r},0)$ is given by Eq.~\eqref{c_ini}, the net shift is calculated as
\begin{align}
\delta r_\alpha = X_\alpha(\bm{x}_\mathrm{ini}) + \frac{\sigma^2}{2} \frac{\partial^2 X_\alpha(\bm{x}_\mathrm{ini})}{\partial x_\beta \partial x_\beta} + \mathcal{O}(\sigma^4). \label{net_shift}
\end{align}
The detailed derivations of Eqs.~\eqref{net_shift_gen} and \eqref{net_shift} are shown in Appendix~\ref{NS_for_GD}.
The analytical result in Eq.~\eqref{net_shift} is plotted as black curves in Fig.~\ref{fig4}, which well reproduce the numerical results.

\section{Summary}

In the present paper, we focus on the dynamics of convection and diffusion, especially on the cooperative effects of them.
The numerical calculations based on the convection-diffusion equation revealed that the diffusion becomes anomalous and the net shift occurs under the reciprocal flow.
Such anomalous phenomena caused by cooperations of the convection and diffusion are understood by the discrete Fokker-Planck equation which is derived from the Langevin dynamics.
Our results are bounded by the condition that the flow is reciprocal and with no vortex, and thus the extension for the case under arbitrary flow profiles remains as future work.

\begin{acknowledgments}
The authors acknowledge Yutaka Abe, Akiko Kaneko, Tadashi Watanabe, Katsuhiro Nishinari, Satoshi Matsumoto, Koji Hasegawa, and Sakurako Tanida for helpful discussion.
This work was supported by JSPS KAKENHI Grant Nos.~JP19K03765, JP19J00365, JP20K14370, JP20H02712, and JP21H01004, and also the Cooperative Research Program of ``Network Joint Research Center for Materials and Devices: Dynamic Alliance for Open Innovation Bridging Human, Environment and Materials'' (Nos.~20211014 and 20214004).
This work was also supported by JSPS and PAN under the Japan-Poland Research Cooperative Program (No.~JPJSBP120204602).
\end{acknowledgments}

\appendix

\section{Flow field\label{FF}}

We consider a flow profile $\bm{v}$ in a two-dimensional system, which satisfies the Navier-Stokes equation~\cite{Landau}:
\begin{align}
\rho \left ( \frac{\partial \bm{v}(\bm{x}, t)}{\partial t} + (\bm{v}(\bm{x}, t) \cdot \nabla) \bm{v}(\bm{x}, t) \right ) \nonumber \\
	= - \nabla p(\bm{x}, t) + \eta \nabla^2 \bm{v}(\bm{x}, t) \label{NS}
\end{align}
with incompressible condition:
\begin{align}
\nabla \cdot \bm{v}(\bm{x}, t) = 0. \label{incompressible}
\end{align}
Here, $\rho$ is the density, $p$ is the pressure, and $\eta$ is the viscosity.

To obtain the solution of Eqs.~\eqref{NS} and \eqref{incompressible}, the velocity potential $\Phi$ is introduced as follows:
\begin{align}
\bm{v}(\bm{x}, t) &= \nabla \Phi(\bm{x}, t), \\
\Phi(\bm{x}, t) &= \Psi(\bm{x}) \cos \omega t.
\end{align}
If $\Psi(\bm{x})$ satisfies the Laplace equation:
\begin{align}
\nabla^2 \Psi(\bm{x}) = 0,
\end{align}
then Eq.~\eqref{incompressible} holds.
In a two-dimensional system, $\Psi(\bm{x})$ satisfies
\begin{align}
&\Psi(\bm{x}) \nonumber \\
&= a_0 + \sum_{n=1}^{\infty} \left ( a_n r^n \cos n (\theta - \theta_n) + \frac{b_n}{r^n} \cos n (\theta - \varphi_n) \right ),
\end{align}
in polar coordinates $(r, \theta)$.
We set $a_0=0$, since the term $a_0$ does not affect the flow.
The term $a_1 r \cos (\theta - \theta_1)$ indicates the shift in the direction of $\theta_1$, which does not change the configuration of the fluid elements.
When the velocity potential is written as $\Phi = a_1 r \cos (\theta - \theta_1) \cos \omega t$, we can take a co-moving frame with the fluid, and thus the flow does not affect the diffusion dynamics.
Therefore, we consider $a_n$ with $n \geq 2$.
Moreover, the term $b_n \cos n (\theta - \varphi_n) / r^n$ diverges at $r=0$, and thus we set $b_n = 0$ for $\forall n$.
Thus, we have 
\begin{align}
\Psi(\bm{x}) = \sum_{n=2}^{\infty} a_n r^n \cos n (\theta - \theta_n).
\end{align}

Next, we check whether the obtained $\Phi$ satisfies Eq.~\eqref{NS}, which is rewritten as
\begin{align}
\rho \frac{\partial \nabla \Phi(\bm{x}, t)}{\partial t} = - \nabla p(\bm{x}, t). \label{NS_mod}
\end{align}
Here, we adopt the low-Reynolds-number limit, i.e., the nonlinear term in Eq.~\eqref{NS} can be neglected.
If we set the pressure field as
\begin{align}
p(\bm{x},t) = \rho \omega \Psi(\bm{x}) \sin \omega t,
\end{align}
the flow profile 
\begin{align}
\bm{v}(\bm{x},t) = \nabla \Psi(\bm{x}) \cos \omega t
\end{align}
satisfies Eqs.~\eqref{NS} and \eqref{incompressible}.

In our numerical calculation and theoretical analysis, a single mode of $n$, i.e.,
\begin{align}
\Psi^{(n)}(\bm{x}) = \frac{\varepsilon}{n} r^n \cos n \theta,
\end{align}
is considered.
Here, $a_m$ is set to be $\varepsilon/n$ for $(m = n)$ and $0$ for $(m \neq n)$, and $\theta_n$ is set to be 0 for simplicity.

\section{Confirmation of the analytical results by numerical simulations\label{Confirm}}

\renewcommand{\thefigure}{A\arabic{figure}}

We numerically calculated the moments $X$, $S$, and $Y$ based on Eq.~\eqref{Langevin}.
Since they are defined as the ensemble average, trajectories of $10^7$ particles were calculated for each initial position $r_\mathrm{ini}(\cos \theta_\mathrm{ini} \bm{e}_x + \sin \theta_\mathrm{ini} \bm{e}_y)$ and parameter set $(D, \varepsilon)$.
The trajectories of the test particles were integrated using the Heun method (second-order Runge-Kutta method) with the time step of $2\pi \times 10^{-3}$, i.e., $10^{3}$ steps for one period.
To generate the Gaussian noise, we adopted the Box-M\"{u}ller method~\cite{Numerical_Recipes}.

\begin{figure}
	\centering
	\includegraphics{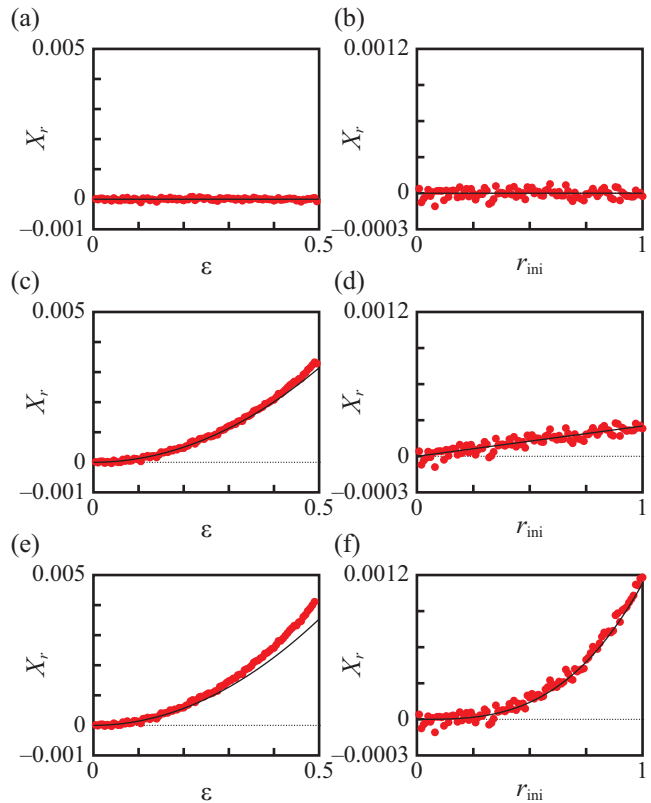}
	\caption{Plots of $X_r$ depending on (a,c,e) $\varepsilon$ and (b,d,f) $r_\mathrm{ini}$ for the mode $n$ ((a,b) $n = 2$, (c,d) 3, and (e,f) 4).
	The thermal diffusion coefficient $D$ and initial angle $\theta_\mathrm{ini}$ were set to be $D= 10^{-3}$ and $\theta_\mathrm{ini} = 0$, respectively.
	The parameter was fixed as $r_\mathrm{ini} = 0.5$ for the left panels, and as $\varepsilon = 0.1$ for the right panels.
	The black solid curves show the theoretical results, which are given in Eq.~\eqref{xr}.}
	\label{figA6}
\end{figure}

The first moments $X_r$ and $X_\theta$, which are in the directions of $r$ and $\theta$, respectively, are defined as
\begin{align}
X_r =& X_x \cos \theta + X_y \sin \theta, \\
X_\theta =& -X_x \sin \theta + X_y \cos \theta
\end{align}
and calculated as in Eqs.~\eqref{xr} and \eqref{xtheta} under the reciprocal flow.
In Fig.~\ref{figA6}, the first moment $X_r$ against the amplitude $\varepsilon$ or the radius of the initial position $r_\mathrm{ini}$ is plotted.
The plots show the numerical results, while the black curves show the theoretical results.
The theoretical curve well predicts the numerical results for smaller $\varepsilon$, because the higher-order terms of $\varepsilon$ do not affect so much in the case with smaller $\varepsilon$.
We also numerically confirmed that $X_\theta$ was almost zero (data not shown), which consists with the theoretical results.

\begin{figure}
	\centering
	\includegraphics{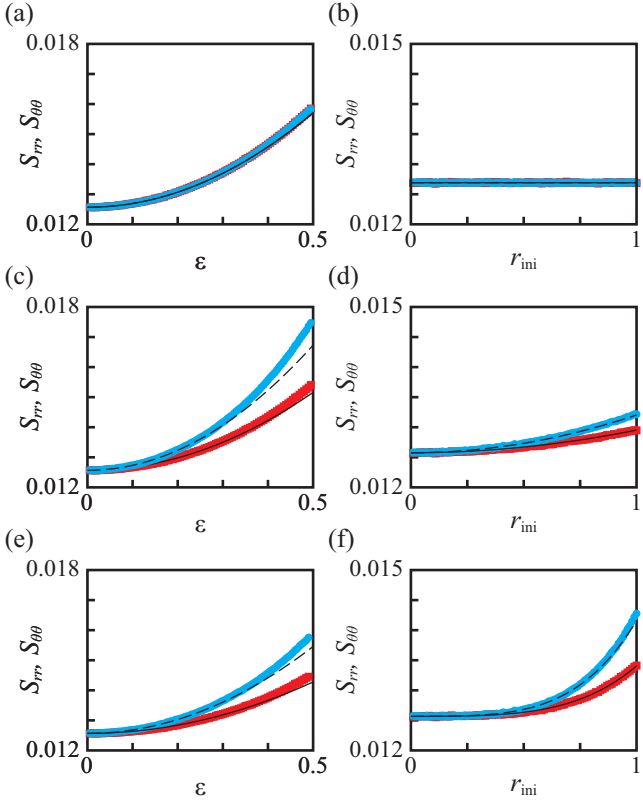}
	\caption{Plots of $S_{rr}$ and $S_{\theta\theta}$ depending on (a,c,e) $\varepsilon$ and (b,d,f) $r_\mathrm{ini}$ for the mode $n$ ((a,b) $n = 2$, (c,d) 3, and (e,f) 4).
	The red and cyan points show $S_{rr}$ and $S_{\theta\theta}$, respectively.
	The thermal diffusion coefficient $D$ and initial angle $\theta_\mathrm{ini}$ were set to be $D= 10^{-3}$ and $\theta_\mathrm{ini} = 0$, respectively.
	The parameter was fixed as $r_\mathrm{ini} = 0.5$ for the left panels, and as $\varepsilon = 0.1$ for the right panels.
	The black solid and dashed curves show the theoretical results for $S_{rr}$ and $S_{\theta\theta}$, which are given in Eqs.~\eqref{srr} and \eqref{stt}, respectively.}
	\label{figA7}
\end{figure}

The second moments $S_{rr}$ and $S_{\theta \theta}$ in the direction of $r$ and $\theta$, and the nondiagonal component $S_{r\theta}$ are defined as
\begin{align}
S_{rr} =& S_{xx} \cos^2\theta + 2S_{xy} \cos \theta \sin \theta + S_{yy} \sin^2\theta, \\
S_{r \theta} =& -S_{xx} \sin \theta \cos \theta + S_{xy} \left( \cos^2 \theta - \sin^2 \theta \right) \nonumber \\
& + S_{yy} \sin \theta \cos \theta, \\
S_{\theta \theta} =& S_{xx} \sin^2\theta - 2S_{xy} \cos \theta \sin \theta + S_{yy} \cos^2\theta,
\end{align}
and they are calculated as:
\begin{align}
S_{rr} =& 2DT + n(n-1) \frac{D\varepsilon^2 T}{\omega^2} r^{2n-4} \nonumber \\
&+ 6(n-1)^2(n-2)^2 \frac{D^2 \varepsilon^2 T^2}{\omega^2} r^{2n-6}, \label{srr}\\
S_{r \theta} =& 0, \\
S_{\theta \theta} =& 2DT + (3n-4)(n-1) \frac{D\varepsilon^2 T}{\omega^2} r^{2n-4} \nonumber \\
&+ 6(n-1)^2(n-2)^2 \frac{D^2 \varepsilon^2 T^2}{\omega^2} r^{2n-6} \label{stt}
\end{align}
under the reciprocal flow.
In Fig.~\ref{figA7}, the second moments, $S_{rr}$ and $S_{\theta\theta}$, against the amplitude $\varepsilon$ or the radius of the initial position $r_\mathrm{ini}$ are plotted.
The numerical results, which are exhibited by red and cyan points, show good correspondence with the theoretical results indicated by the black curves.
We also numerically confirmed that $S_{r \theta}$ was almost zero (data not shown), which consists with the theoretical results.

\begin{figure}
	\centering
	\includegraphics{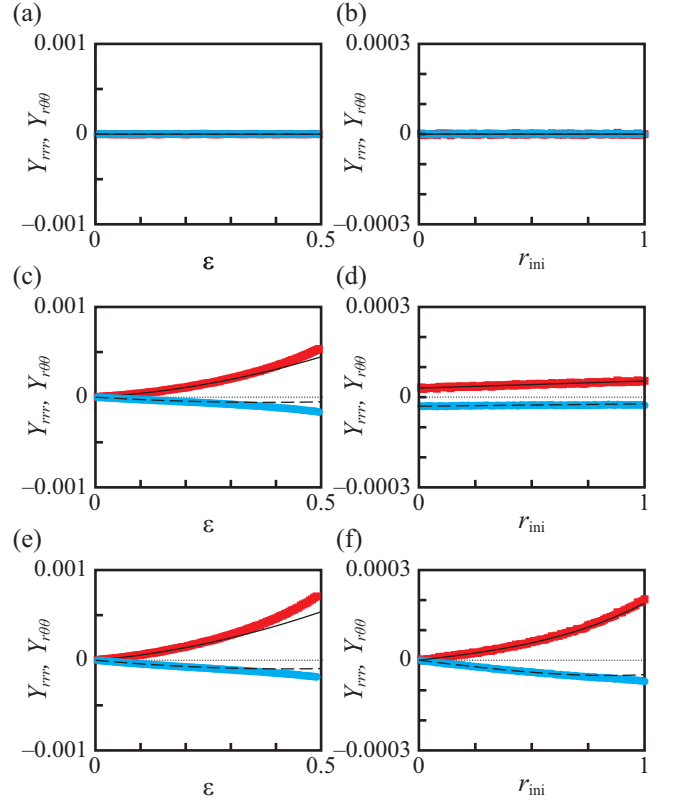}
	\caption{Plots of $Y_{rrr}$ and $Y_{r\theta\theta}$ depending on (a,c,e) $\varepsilon$ and (b,d,f) $r_\mathrm{ini}$ for the the mode $n$ ((a,b) $n = 2$, (c,d) 3, and (e,f) 4).
	The red and cyan points show $Y_{rrr}$ and $Y_{r\theta\theta}$, respectively.
	The thermal diffusion coefficient $D$ and initial angle $\theta_\mathrm{ini}$ were set to be $D= 10^{-3}$ and $\theta_\mathrm{ini} = 0$, respectively.
	The parameter was fixed as $r_\mathrm{ini} = 0.5$ for left panels, and as $\varepsilon = 0.1$ for right panels.
	The black solid and dashed curves show the theoretical results for $Y_{rrr}$ and $Y_{r\theta\theta}$, which are given in Eqs.~\eqref{yrrr} and \eqref{yrtt}.}
	\label{figA8}
\end{figure}

\begin{figure}
	\centering
	\includegraphics{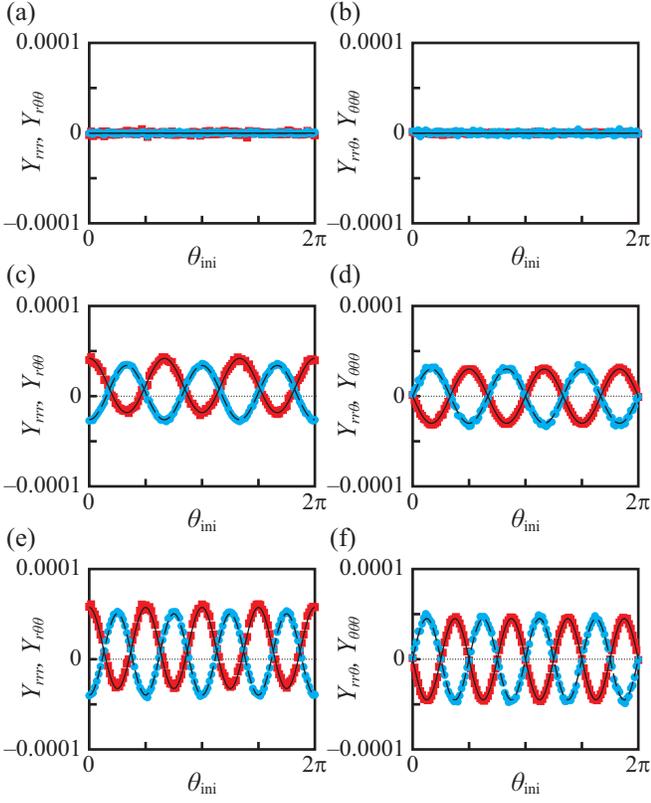}
	\caption{Plots of (a,c,e) $Y_{rrr}$ and $Y_{r\theta\theta}$, and (b,d,f) $Y_{rr\theta}$ and $Y_{\theta\theta\theta}$ depending on $\theta$ for the mode $n$ ((a,b) $n = 2$, (c,d) 3, and (e,f) 4).
	The red points show (a,c,e) $Y_{rrr}$ and (b,d,f) $Y_{rr\theta}$, and the cyan points show (a,c,e) $Y_{r\theta\theta}$ and (b,d,f) $Y_{\theta\theta\theta}$.
	The thermal diffusion coefficient $D$, amplitude of the flow field $\varepsilon$, and initial radius $r_\mathrm{ini}$ were set to be $D= 10^{-3}$, $\varepsilon = 0.1$, and $r_\mathrm{ini} = 0.5$, respectively.
	The black solid curves show the theoretical results for (a,c,e) $Y_{rrr}$ and (b,d,f) $Y_{rr\theta}$, and the black dashed curves show the theoretical results for (a,c,e) $Y_{r\theta\theta}$ and (b,d,f) $Y_{\theta\theta\theta}$, which are given in Eqs.~\eqref{yrrr}-\eqref{yttt}.}
	\label{figA9}
\end{figure}

The third moments $Y_{rrr}$, $Y_{rr\theta}$, $Y_{r\theta\theta}$, and $Y_{\theta\theta\theta}$ are defined as:
\begin{align}
Y_{rrr} =& Y_{xxx} \cos^3 \theta + 3Y_{xxy} \cos^2 \theta \sin \theta \nonumber \\
& + 3Y_{xyy} \cos \theta \sin^2 \theta + Y_{yyy} \sin^3 \theta, \\
Y_{rr \theta} =& -Y_{xxx} \cos^2\theta \sin\theta + Y_{xxy} \left( \cos^3\theta - 2\cos \theta \sin^2 \theta\right) \nonumber \\
& + Y_{xyy} \left(2 \cos^2 \theta \sin \theta - \sin^3 \theta \right) + Y_{yyy} \cos \theta \sin^2 \theta, \\
Y_{r \theta \theta} =& Y_{xxx} \cos \theta \sin^2 \theta + Y_{xxy} \left( \sin^3 \theta - 2 \cos^2 \theta \sin \theta \right) \nonumber \\
& + Y_{xyy} \left( \cos^3 \theta - 2 \cos \theta \sin^2 \theta \right) + Y_{yyy} \cos^2 \theta \sin \theta, \\
Y_{\theta \theta \theta} =& -Y_{xxx} \sin^3 \theta + 3Y_{xxy} \cos \theta \sin^2 \theta \nonumber \\
& - 3Y_{xyy} \cos^2 \theta \sin\theta + Y_{yyy} \cos^3 \theta,
\end{align}
and are calculated as:
\begin{align}
Y_{rrr} =& 24(n-1)(n-2) \frac{D^2 \varepsilon T}{\omega^2} r^{n-3} \cos n \theta \nonumber \\
& + 6(n-1)(n-2)(2n-1) \frac{D^2\varepsilon^2 T^2}{\omega^2} r^{2n-5}, \label{yrrr} \\
Y_{rr \theta} =& -24(n-1)(n-2) \frac{D^2 \varepsilon T}{\omega^2} r^{n-3} \sin n \theta, \label{yrrt} \\
Y_{r \theta \theta} =& - 24(n-1)(n-2) \frac{D^2 \varepsilon T}{\omega^2} r^{n-3} \cos n \theta \nonumber \\
& + 2(n-1)(n-2)(4n-7) \frac{D^2\varepsilon^2 T^2}{\omega^2} r^{2n-5}, \label{yrtt} \\
Y_{\theta \theta \theta} =& 24(n-1)(n-2) \frac{D^2 \varepsilon T}{\omega^2} r^{n-3} \sin n \theta \label{yttt} 
\end{align}
under the reciprocal flow.
It is noted that the third moments depend on not only $r$ but also $\theta$.
In Fig.~\ref{figA8}, the third moments $Y_{rrr}$ and $Y_{r\theta\theta}$ against the amplitude $\varepsilon$ or the radius of the initial position $r_\mathrm{ini}$ are plotted.
In Fig.~\ref{figA9}, the dependence of $Y_{rrr}$, $Y_{r\theta\theta}$, $Y_{rr\theta}$, and $Y_{\theta\theta\theta}$ on $\theta_\mathrm{ini}$ are plotted.
The theoretical results drawn as black curves well reproduce the numerical results plotted as red and cyan points.

\section{Net shift for the Gaussian distribution\label{NS_for_GD}}

In this section, the net shift of the Gaussian distribution in Eq.~\eqref{c_ini} after one period of the reciprocal flow is calculated.

The first moment at $t=0$ is given by
\begin{align}
\overline{x}_{\alpha} = \int x_\alpha c(\bm{x},0) d\bm{x},
\end{align}
where we assume that $c(\bm{x},0)$ is normalized.
To consider the concentration profile at $t=T$, the transition probability $Q(\bm{x}, \bm{x}')$ from $\bm{x}'$ at $t=0$ to $\bm{x}$ at $t = T$ is introduced as:
\begin{align}
c(\bm{x},T) = \int Q(\bm{x}, \bm{x}') c(\bm{x}',0) d\bm{x}',
\end{align}
which satisfies the law of mass conservation:
\begin{align}
\int Q(\bm{x}, \bm{x}') d\bm{x} = 1.
\end{align}
Using the transition probability, the first moment $X_{\alpha}(\bm{x})$ can be represented as 
\begin{align}
X_{\alpha}(\bm{x}) = \int (x'_\alpha - x_\alpha) Q(\bm{x}', \bm{x}) d\bm{x}'.
\end{align}

Then, the expression of the net shift $\delta \bm{r}$ for an arbitrary concentration profile can be derived as follows:
\begin{align}
\delta r_\alpha 
=& \int r_\alpha c(\bm{r},T) d\bm{r} - \overline{x}_\alpha \nonumber \\
=& \int r_\alpha \int Q(\bm{r},\bm{r}') c(\bm{r}',0) d\bm{r}' d\bm{r} - \overline{x}_\alpha \nonumber \\
=& \int \left [ \int (r_\alpha - r'_\alpha) Q(\bm{r},\bm{r}') d\bm{r} \right . \nonumber \\
& \left . \qquad + r'_\alpha \int Q(\bm{r},\bm{r}') d\bm{r} \right ] c(\bm{r}',0) d\bm{r}' - \overline{x}_\alpha \nonumber \\
=& \int \left [ X_\alpha(\bm{r}') + r'_\alpha \right ] c(\bm{r}',0) d\bm{r}' - \overline{x}_\alpha \nonumber \\
=& \int X_\alpha(\bm{r}') c(\bm{r}',0) d\bm{r}'.
\end{align}
When the Gaussian distribution in Eq.~\eqref{c_ini} is considered, the net shift $\delta \bm{r}$ is calculated as:
\begin{align}
&\delta r_\alpha \nonumber \\
&= \int X_\alpha(\overline{\bm{x}} + \bm{\rho}) c(\overline{\bm{x}} + \bm{\rho},0) d\bm{\rho} \nonumber \\
&= \int \left [ X_\alpha(\overline{\bm{x}}) + \frac{\partial X_\alpha(\overline{\bm{x}})}{\partial x_\beta} \rho_\beta + \frac{1}{2} \frac{\partial^2 X_\alpha(\overline{\bm{x}})}{\partial x_\beta \partial x_\gamma} \rho_\beta \rho_\gamma + \mathcal{O}(|\bm{\rho}|^3) \right ] \nonumber \\
& \qquad \times c(\overline{\bm{x}} + \bm{\rho},0) d\bm{\rho} \nonumber \\
&= X_\alpha(\overline{\bm{x}}) + \frac{\sigma^2}{2} \frac{\partial^2 X_\alpha(\overline{\bm{x}})}{\partial x_\beta \partial x_\beta} + \mathcal{O}(\sigma^4).
\end{align}

\end{document}